\documentclass[%
 aip,
rsi,%
 amsmath,amssymb,
 reprint,%
]{revtex4-1}

\usepackage[dvipsnames]{xcolor}
\usepackage{graphicx}
\usepackage{dcolumn}
\usepackage{bm}
\usepackage{lipsum}

\usepackage{dcolumn}
\usepackage{epstopdf}
\usepackage{xcolor}
\usepackage{natbib}
\usepackage{subcaption} 
\usepackage{float}
\usepackage{tabularx}
\usepackage{mathrsfs}
\usepackage{upgreek}
\usepackage[a]{esvect}
\setcitestyle{square}
\usepackage[toc,page]{appendix}
\usepackage[font={small,normalfont}]{caption}
\usepackage[colorlinks=true, allcolors=blue]{hyperref}
\usepackage[normalem]{ulem}
\usepackage[capitalise]{cleveref}

\begin{document}


\title{Phase-controlled direct laser acceleration enabled by longitudinal variation of the laser-driven quasi-static plasma magnetic field}

\author{R. Bhakta}
\affiliation{Department of Mechanical and Aerospace Engineering, University of California San Diego, La Jolla, CA 92093}

\author{I-L. Yeh}
\affiliation{Department of Physics, University of California San Diego, La Jolla, CA 92093}

\author{K. Tangtartharakul}
\affiliation{Department of Mechanical and Aerospace Engineering, University of California San Diego, La Jolla, CA 92093}

\author{L. Willingale}
\affiliation{Gérard Mourou Center for Ultrafast Optical Science, University of Michigan, Ann Arbor, Michigan 48109}

\author{A. Arefiev}
\email[]{aarefiev@ucsd.edu}

\affiliation{Department of Mechanical and Aerospace Engineering, University of California San Diego, La Jolla, CA 92093}

\affiliation{Center for Energy Research, University of California San Diego, La Jolla, CA 92093}

\date{\today}

\begin{abstract}

Direct laser acceleration (DLA) enables energy transfer from an ultra-high-intensity laser to plasma electrons and underpins many laser-driven particle and radiation-source concepts. A laser-driven azimuthal plasma magnetic field is a key player in this process: it confines energetic electrons, induces betatron oscillations, and makes possible a resonant interaction between the betatron motion and the laser field. While this betatron resonance can enhance electron energy gain, the gain itself generally drives frequency detuning and promotes largely reversible energy exchange that limits net acceleration. Here we show, using a test-electron model with prescribed fields, that a slow longitudinal increase of the quasi-static plasma magnetic field qualitatively changes DLA by introducing hysteresis in the ratio of the betatron frequency to the laser frequency experienced by the electron, so that this ratio depends on the prior evolution of the electron even at the same energy. This hysteresis enables phase control of the electron–laser energy exchange and suppresses the usual reversibility of DLA, allowing electrons to retain the acquired energy and sustain energy gain without intermittent losses.

\end{abstract}

\maketitle


\section{Introduction} \label{sec: intro}

High-power high-intensity lasers have found a wide range of applications, including laser-driven particle~\cite{umstadter.pop.2001,daido.2012} and radiation sources~\cite{yun.rmpp.2025}. Often, energy transfer from the laser field to the target is a critical step. Specifically, due to the high laser intensity, the target is a plasma, so the first step is the energy transfer from the laser field to plasma electrons. The underlying mechanism at relativistic laser intensities is often referred to as direct laser acceleration or, for short, DLA\cite{gahn.prl.1999,arefiev.jpp.2015, arefiev.pop.2016}. Recently, DLA has attracted additional interest as multi-petawatt laser facilities such as ELI~\cite{weber2017p3, gales2018eli-np, lureau.hplse.2020,Cernaianu2025,Doria.JI.2020.ELINP}, ZEUS~\cite{Nees.zeus.2020,willingale.zeus.2023}, and CoReLS~\cite{yoon.optica.2021} became fully operational.


The basic concept of DLA can be understood using a simple model involving a single electron and a plane electromagnetic wave. The transverse electric field accelerates the electron transversely while transferring energy from the wave. Once the electron becomes relativistic, the transverse magnetic field provides a strong push in the forward direction and redirects the acquired energy toward forward motion. As a result, the wave, representing the laser, produces an energetic forward-moving electron. When the laser interacts with an electron population rather than a single electron, the same mechanism leads to the generation of an energetic electron beam~\cite{pukhov.2003,arefiev.jpp.2015,macchi.2013}.


How to effectively leverage DLA remains an active area of research~\cite{rosmej.ppcf.2020,kemp.pop.2020,hussein.njp.2021,Tang.NJP.2024.Focusing, babjak.prl.2024,valenta.pre.2024,rosmej.hplse.2025,cohen.sciadv.2024,tang.pop.2025}. The oscillating nature of the laser field presents specific challenges. Since the wavefronts propagate forward faster than the electron, the electron continuously slips in phase. As a result, the energy-gain stage in this picture lasts only for a limited distance. Moreover, phase slippage makes the energy exchange largely reversible. In addition, as the magnetic field redirects the motion toward the forward direction, the rate of energy transfer decreases because the transverse component of the electron velocity becomes progressively smaller. 

The plasma fields that arise during laser propagation introduce additional physics that has been explored as a way to mitigate these difficulties. During laser propagation through a plasma, the laser beam expels some electrons transversely while pushing other electrons forward, generating a field configuration with quasi-static radial electric and azimuthal magnetic fields~\cite{arefiev.pop.2016,stark.prl.2016,jansen.ppcf.2018,jirka.njp.2020,wang.pop.2020,babjak.prl.2024}. These plasma fields deflect laser-accelerated electrons toward the laser axis, providing transverse confinement. The oscillations caused by these deflections are called betatron oscillations and they are a key physics element. Betatron oscillations can mitigate the negative effect of phase slippage by enabling a resonant interaction, often referred to as the betatron resonance~\cite{pukhov.pop.1999,khudik.pop.2016,yeh.pop.2025}. If the betatron oscillation frequency matches the oscillation frequency of the laser field experienced by the electron, then the transverse electron velocity can remain antiparallel to the laser electric field. This allows the electron to continue gaining energy despite phase slippage, because the phase slippage no longer controls the relative orientation of the transverse velocity and the laser field~\cite{arefiev.pop.2016,gong.pre.2020}.

The betatron resonance can significantly boost the energies attained by plasma electrons. However, the relevant frequencies, in general, have different energy dependence~\cite{khudik.pop.2016,arefiev.pop.2024,yeh.pop.2025}. As a result, the energy gain itself becomes detrimental by driving the system away from resonance~\cite{yeh.pop.2025}. It turns out that the frequency detuning can be mitigated to allow further energy gain. The mitigating factor is somewhat counterintuitive: it is the superluminal phase velocity. It has been shown that a certain degree of superluminosity can keep the frequency ratio close to resonance as the electron energy grows~\cite{khudik.pop.2016,arefiev.pop.2024,yeh.pop.2025,tangtartharakul.njp.2025}. Simulations and analytical models nevertheless indicate that the energy exchange during DLA is typically fully reversible~\cite{khudik.pop.2016,tangtartharakul.njp.2025}. This reversibility can be managed by decoupling energetic electrons from the laser on a time scale shorter than the deceleration time. However, this raises a key question of whether reversibility is an intrinsic feature of DLA.

\begin{figure}[t]
    \begin{center} \includegraphics[width=1\columnwidth,clip]{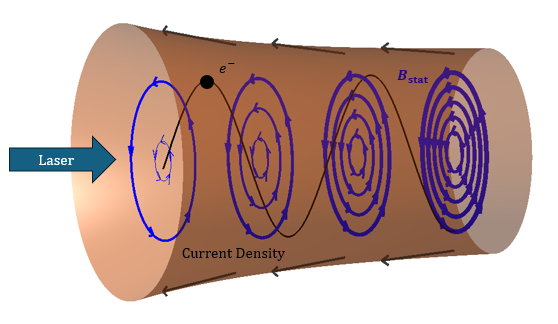} \caption{\label{fig:schematic} Schematic of the plasma current filament used in the test-electron model, with the return current that typically flows on the periphery~\cite{gong.pre.2020} not shown. The filament has both longitudinal and radial current density components. The brown surface is an isosurface of the current-density magnitude $|\bm{j}|$. The black arrows indicate the local direction of the current-density vector $\bm{j}$. The blue circles show the azimuthal magnetic field produced by the filament, with several cross sections illustrating its variation along the laser propagation direction. The black curve is an example of a flat electron trajectory in this field, with a characteristic forward drift.}
    \end{center}
\end{figure}


In this work, we examine the role of a longitudinally nonuniform plasma magnetic field that can naturally arise in laser--plasma interactions. To clearly isolate the impact of this nonuniformity on the electron dynamics, we use a test-electron model in which the laser and plasma fields are prescribed and there is no feedback from the electron motion. Such models are commonly used to study DLA~\cite{arefiev.pop.2016,khudik.pop.2016,arefiev.pre.2020,gong.pre.2020,babjak.prl.2024}, often with the simplifying assumption that the plasma fields are longitudinally uniform. We find that even a slow longitudinal increase of the current density and, therefore, of the plasma magnetic-field strength along the electron trajectory can have a profound impact. Specifically, the frequency ratio that controls the resonant interaction develops hysteresis, in the sense that its value at a given electron energy becomes dependent on the prior motion of the electron. As a result, the energy exchange that is typically reversible in DLA is no longer reversible. We show that this hysteresis can substantially increase the attainable electron energy. We also show that it leads to two important outcomes: energy retention and steady energy gain without intermittent losses.

The remainder of this paper consists of five sections. \Cref{sec-tes_electron_model} introduces the test-electron model with a longitudinally varying plasma magnetic field that we use in this work. In \cref{sec-alphaConst}, we revisit the case of a longitudinally uniform plasma magnetic field with a superluminal phase velocity to establish a clear baseline for comparison. In \cref{sec-linear}, we introduce a linear dependence of the current density along the electron trajectory and show that this gives us the ability to regulate the evolution of the offset between the laser phase and the betatron phase, thereby substantially prolonging the energy gain. In \cref{sec: regulation}, we show that, by regulating the phase offset, one can achieve two distinct outcomes introduced earlier: energy retention, and steady energy gain without intermittent losses. Finally, \cref{sec: summary} provides a brief summary of our main results.


\section{Test-electron model with a longitudinally varying azimuthal plasma magnetic field} \label{sec-tes_electron_model}

We analyze the electron dynamics using a test-electron model in which the laser and plasma fields are prescribed rather than obtained self-consistently. This approach is appropriate for examining energetic-electron dynamics because such electrons typically constitute only a small fraction of the total population and therefore do not strongly modify the fields. Test-electron models have played an important role in studies of DLA~\cite{arefiev.pop.2016,khudik.pop.2016,jirka.njp.2020,arefiev.pre.2020,gong.pre.2020,babjak.prl.2024,valenta.pre.2024}. These models differ primarily in how the laser and plasma fields are represented and in whether radiation reaction is included.

Particle-in-cell (PIC) simulations provide a self-consistent description of the laser-plasma interaction, so they are often used to inform the setup for test-electron models~\cite{arefiev.pop.2016,gong.pre.2020}. They are particularly useful for providing insights into the configuration of the plasma fields. Specifically, PIC simulations show that an ultraintense laser beam tends to create a channel in the plasma with radial electric and azimuthal magnetic fields that vary slowly in time~\cite{arefiev.pop.2016,stark.prl.2016,jansen.ppcf.2018,jirka.njp.2020,wang.pop.2020,babjak.prl.2024}. This is why test-electron models often include these fields. For simplicity, these fields are often represented as static and uniform along the laser propagation direction. The assumption of longitudinal uniformity makes the electron dynamics more tractable, but it also removes the ability to examine how longitudinal variations influence the electron dynamics. The model presented in this section is designed to address this limitation. Since PIC simulations do show that these variations naturally arise as the laser propagates through the plasma~\cite{stark.prl.2016,gong.pre.2020}, the introduction of a controlled longitudinal dependence presented here is a natural next step in the development of test-electron models for DLA.

In our test-electron model, the laser is represented by a plane electromagnetic wave, whereas the plasma field is taken to be only a static azimuthal magnetic field that varies along the laser propagation direction. Such a plasma magnetic field can be maintained by a cylindrically symmetric current filament that is schematically shown in \cref{fig:schematic}. Our model extends the model introduced in Ref.~[\onlinecite{yeh.pop.2025}] by introducing the longitudinal variation of the magnetic field. The radial electric field is intentionally not included. This simplifies the analysis and enables us to pinpoint the exact role of the longitudinal nonuniformity of the plasma magnetic field. The simplification is also motivated by self-consistent PIC simulations in which the azimuthal magnetic field can greatly exceed the accompanying radial electric field~\cite{jansen.ppcf.2018}. We adopt the notations and terminology used in Ref.~[\onlinecite{yeh.pop.2025}] to facilitate comparison with the longitudinally uniform case treated in detail there.

The time evolution of the electron momentum $\bm{p}$ and position $\bm{r}$ are described by
\begin{eqnarray}
&& \frac{d \bm{p}}{d t} = - |e| \bm{E} - \frac{|e|}{\gamma m_e c} \left[ \bm{p} \times \bm{B} \right], \label{dpdt} \\
&& \frac{d \bm{r}}{d t} = \frac{\bm{p}}{\gamma m_e}, \label{drdt} 
\end{eqnarray}
where $m_e$ is the electron mass and 
\begin{equation} \label{gamma}
    \gamma = \sqrt{1 + \bm{p}^2/ m_e^2 c^2}
\end{equation}
is the relativistic factor. The electric and magnetic fields are a superposition of the oscillating laser fields and a static plasma magnetic field, denoted by the subscripts ``laser'' and ``stat'': $\bm{E} = \bm{E}_{\rm{laser}}$ and $\bm{B} = \bm{B}_{\rm{laser}} + \bm{B}_{\rm{stat}}$. 

The laser fields are represented as a plane, linearly polarized electromagnetic wave with a phase velocity $v_{ph}$ and frequency $\omega_0$  propagating in the positive $x$-direction. The fields of the laser are then
\begin{eqnarray}
     &&\bm{E}_{\rm{laser}} = \bm{e}_y E_0 \cos(\xi), \label{E-laser}\\
     &&\bm{B}_{\rm{laser}} = \bm{e}_z \frac{E_0}{u}  \cos(\xi), \label{B-laser}
\end{eqnarray}
where $E_0$ is a constant laser amplitude,  
\begin{equation}
    u \equiv v_{ph} / c
\end{equation}
is the normalized phase velocity, and
\begin{equation} \label{xi}
    \xi = \omega_0 t - \omega_0 x/v_{ph} + \xi_* = \omega_0 t - \omega_0 x/ u c + \xi_*
\end{equation}
is the laser phase, with $\xi_*$ being the initial phase at the electron location $x=0$ at $t=0$. In what follows, we will characterize the laser strength using the normalized amplitude
\begin{equation}
    a_0 \equiv |e| E_0 / m_e c \omega_0.
\end{equation}

In our model, the static plasma magnetic field is generated by a cylindrically symmetric plasma current filament with current density $\bm{j}$. The longitudinal component $j_x (x)$ varies slowly along the laser propagation direction, while remaining uniform across each transverse cross section. In contrast to earlier test-electron models, the current filament in our formulation also has a radial component. This component ensures charge conservation by keeping the current density divergence-free, $\nabla \cdot \bm{j} = 0$. For a prescribed longitudinal component $j_x (x)$, the radial component is 
\begin{equation}
    j_r = - \frac{r}{2} \frac{d j_x}{dx}.
\end{equation}
As a result, the magnitude of the current density becomes radius-dependent, with
\begin{equation}
    |\bm{j}| = \sqrt{ j_x^2 (x) + \frac{r^2}{4} \left( \frac{d j_x(x)}{dx} \right)^2}. 
\end{equation} 
\Cref{fig:schematic} shows an example of an isosurface of $| \bm{j} |$, illustrating how it narrows as $|j_x|$ increases along the laser propagation direction. 

The magnetic field produced by the current filament is purely azimuthal, as illustrated in \cref{fig:schematic}. It is given by
\begin{equation}
    \bm{B}_{\rm{stat}} = -  \frac{2 m_e c^2}{|e|} \frac{\alpha(x)}{\lambda_0^2} r \bm{e}_{\phi}.
    \label{B_stat_3D}
\end{equation}
Here
\begin{equation} \label{eq: def alpha}
    \alpha(x) \equiv \pi \lambda_0^2 |j_x (x)|/J_A
\end{equation}
is a dimensionless current parameter, with \(J_A = m_e c^3 / |e|\) being the classical Alfvén current and $\lambda_0 = 2 \pi c / \omega_0$ being the vacuum laser wavelength. We explicitly assume that $j_x < 0$, as is typical for a laser-driven current filament. 

In this study, we restrict attention to flat electron trajectories. These are trajectories in the \((x,y)\)-plane ($z=0$), which contains the axis of the current filament. In this plane, the magnetic field has only one component,
\begin{equation}
    B_{{\rm{stat}}, z} = -  \frac{2 m_e c^2}{|e|} \frac{\alpha(x)}{\lambda_0^2} y.
    \label{B_stat}
\end{equation}
There is no force to push an electron with $\bm{p} = (p_x, p_y,0)$ out of the plane. The laser magnetic field also only has a $z$-component, whereas the laser electric field lies in the $(x,y)$-plane. That is why flat trajectories are possible in this setup. The sign of the plasma magnetic field reverses as the electron crosses the axis, so the resulting force produces the characteristic forward-sliding transverse motion shown in \cref{fig:schematic}, even in the absence of the laser, rather than a closed cyclotron orbit.

In summary, our test-electron model consists of Eqs.~(\ref{dpdt}) and (\ref{drdt}), where the fields are the laser fields given by Eqs.~(\ref{E-laser}) and (\ref{B-laser}), together with the plasma magnetic field given by \cref{B_stat}, which has longitudinal dependence. In addition to electron initial conditions, the model requires three dimensionless inputs: the normalized laser amplitude $a_0$, the normalized phase velocity $u$, and the prescribed longitudinal profile of the normalized longitudinal current density $\alpha(x)$. 


\section{Baseline dynamics for a longitudinally uniform plasma magnetic field} \label{sec-alphaConst}

In this section, we revisit the case of a longitudinally uniform plasma magnetic field with a superluminal laser phase velocity ($u > 1$), first analyzed in Ref.~[\onlinecite{khudik.pop.2016}] and subsequently extended in Ref.~[\onlinecite{yeh.pop.2025}]. The purpose is to establish a clear baseline for comparison once longitudinal variations of the magnetic field are introduced.

In a longitudinally uniform plasma magnetic field, the current parameter $\alpha$ is constant and this longitudinal symmetry gives rise to an integral of motion. The quantity
\begin {equation}
\label{eq:S}
    S = \gamma - u\frac{p_x}{m_e c} +u \alpha \frac{y^2}{\lambda_0^2}
\end {equation}
remains conserved throughout the electron's motion in the combined fields of the laser and the plasma~\cite{yeh.pop.2025}. For given $\alpha$, $u$, and $a_0$, this quantity $S$ uniquely determines the maximum attainable $\gamma$.

\begin{figure}[t]
    \begin{center}
    \includegraphics[width=1\columnwidth,clip]{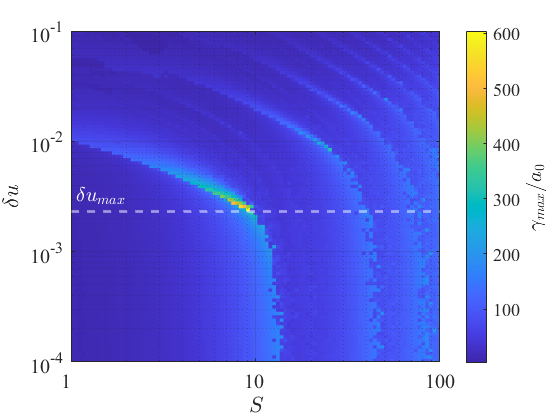}
    \caption{\label{fig:duScan} Maximum $\gamma$, denoted as $\gamma_{\max}$, attained by an electron in a longitudinally uniform plasma magnetic field as a function of the relative superluminosity $\delta u$ and the conserved quantity $S$. The horizontal dashed line marks $\delta u_{\max}$, the value of $\delta u$ that yields the highest $\gamma_{\max}$. The scan is for $a_0 = 8$ and $\alpha = 1.4$.}
    \end{center}
\end{figure}

A key result established in Refs.~[\onlinecite{khudik.pop.2016}] and [\onlinecite{yeh.pop.2025}] is that slight superluminosity of the laser phase velocity is advantageous. Specifically, a properly chosen degree of superluminosity,
\begin{equation}
    \delta u \equiv u - 1 = (v_{ph} - c)/c,
\end{equation}
can strongly enhance the maximum attainable electron energy, even though $\delta u \ll 1$. The $\delta u$ scan in \cref{fig:duScan}, obtained by solving the test-electron model equations from \cref{sec-tes_electron_model} for $a_0 = 8$ and $\alpha = 1.4$, illustrates this effect. These $a_0$ and $\alpha$ are chosen to provide a concrete reference case. The resulting $\gamma_{\max}$, defined as the highest $\gamma$ attained by the electron and calculated as explained below, exhibits a series of distinct bands of enhanced energy gain~\cite{yeh.pop.2025}. Within each band, $\gamma_{\max}$ peaks at a small but finite value of $\delta u$.

In the scan, the electron starts on the axis ($y=0$) with an initial transverse momentum $p_y$ chosen to satisfy the prescribed value of $S$. To account for the sensitivity of the dynamics to the initial laser phase $\xi_*$, each $(\delta u,S)$ pair was simulated for 400 uniformly spaced values $\xi_* = 2\pi k/400$ with $k = 0,1,\dots,399$, thereby sampling one full $2\pi$ interval without double counting $\xi_*=2\pi$ (which is equivalent to $\xi_*=0$). The equations of motion [\cref{dpdt,drdt}] were integrated up to $t\omega_0 = 3\times 10^{5}$ for each phase. For each $(\delta u,S)$ pair, $\gamma_{\max}$ is taken as the largest attained $\gamma$ across the ensemble of phases.


\begin{figure}[t]
    \begin{center}
    \includegraphics[width=1\columnwidth,clip]{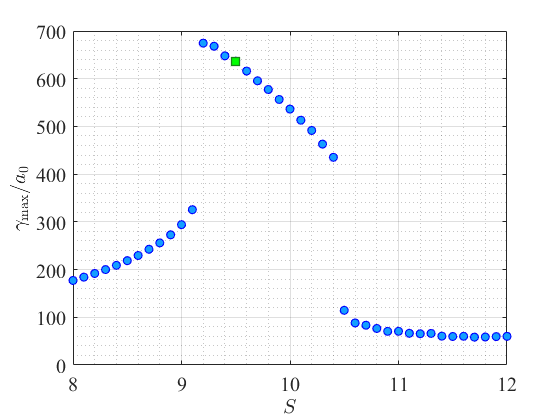}
    \caption{\label{fig:duMax} $\gamma_{\max}$ as a function of $S$ at $\delta u = \delta u_{\max}$, where $\delta u_{\max}$ is the value of $\delta u$ that yields the highest $\gamma_{\max}$ in the scan shown in \cref{fig:duScan}. The green marker highlights the value of $S$ corresponding to the trajectory analyzed in \cref{fig:alphaConstFreq} and \cref{fig:alphaConstPhase}. }
    \end{center}
\end{figure}

In the scan presented in \cref{fig:duScan}, the relative degree of superluminosity that produces the highest $\gamma_{\max}$ is $\delta u_{\max} \approx 2.2 \times 10^{-3}$. The electron attains this $\gamma_{\max}$ within the left-most band of enhanced energy gain. \Cref{fig:duMax} shows $\gamma_{\max}$ as a function of $S$ at this specific value of $\delta u$ to highlight the dependence of $\gamma_{\max}$ on $S$ within the band. Two thresholds in $S$ are visible, delimiting a finite interval over which electrons experience enhanced energy gain. Hereafter, we focus on this specific interval, referring to it simply as the energy-gain band. 

To understand the origin of the energy-gain band, we turn our attention to the electron dynamics. There are two characteristic frequencies that influence the electron energy gain by setting the relative orientation of the transverse electron velocity and the transverse laser electric field. These frequencies are the betatron frequency $\omega_{\beta}$, which describes transverse electron oscillations induced by the plasma magnetic field, and the laser frequency experienced by the electron, $\omega' = d \xi / dt$. The latter is not simply $\omega_0$, because the electron moves along the $x$-axis with a relativistic velocity. Moreover, this velocity oscillates during betatron motion, so the rate at which the electron slips through the laser phase varies over a betatron period. Therefore, a representative measure for the laser field oscillations at the electron location is $\langle \omega' \rangle$, which is $\omega'$ averaged over one betatron oscillation. 

If $\langle \omega' \rangle = \omega_{\beta}$, then the electron experiences a betatron resonance. The resonance leads to a net energy gain with each betatron oscillation. A challenge for sustaining this resonant energy gain is that $\langle \omega' \rangle$ and $\omega_{\beta}$ have different dependencies on $\gamma$. As the electron energy increases, these differing scalings cause the two frequencies to drift apart, leading to detuning. A slight superluminosity can mitigate this effect. For small $\delta u$, the ratio of the two frequencies can be approximated as~\cite{yeh.pop.2025}
\begin{equation}
    \frac{\langle \omega' \rangle}{\omega_{\beta}} \approx \frac{\pi}{\sqrt{2 \alpha}} \left( S \gamma^{-1/2} + 3 \gamma^{1/2} \delta u \right). \label{eq:freq_ratio}
\end{equation}
In the luminal case ($\delta u = 0$), the ratio decreases monotonically as $\gamma$ increases. A slight superluminal phase velocity introduces a non-monotonic dependence in $\gamma$ with a global minimum. If the minimum value of this ratio is close to unity, then the superluminal phase velocity is able to delay the onset of detuning.

\begin{figure}[t]
    \begin{center}
    \includegraphics[width=.87\columnwidth,clip]{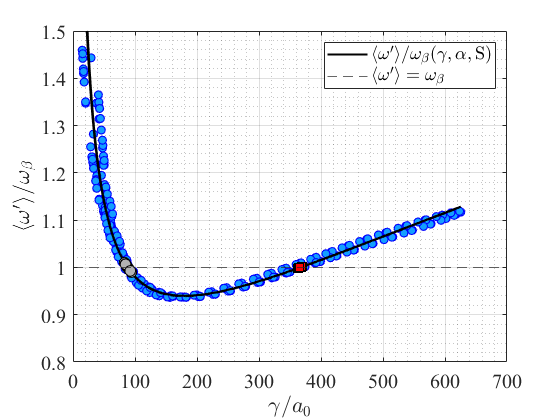}
    \caption{\label{fig:alphaConstFreq} Frequency ratio $\langle \omega' \rangle / \omega_{\beta}$ as a function of $\gamma$ for $S = 9.5$ and $\delta u = \delta u_{\max}$ ($\alpha = 1.4$, $a_0 = 8$). The solid curve is the analytical result given by \cref{eq:freq_ratio} and the blue round markers are the values obtained during numerical integration of the equations of motion. The horizontal dashed line marks the betatron resonance, $\langle \omega' \rangle = \omega_{\beta}$. The round gray and square red markers indicate the values of $\gamma$ at which this resonance condition is satisfied.}
    \end{center}
\end{figure}

The energy-gain band in \cref{fig:duMax} is enabled by the delayed detuning. In this band, $\langle \omega' \rangle / \omega_{\beta}$ depends on $\gamma$ similarly to what is shown in \cref{fig:alphaConstFreq}. This specific example is generated for $S = 9.5$, whose $\gamma_{\max}$ is marked with the green square in \cref{fig:duMax}. The frequency ratio shown in \cref{fig:alphaConstFreq} includes both the analytical expression given by \cref{eq:freq_ratio} and values evaluated numerically when integrating the equations of motion, following the procedure described in Ref.~[\onlinecite{yeh.pop.2025}]. The key feature is that the frequency ratio remains close to the resonance condition $\langle \omega' \rangle = \omega_{\beta}$ over a broad range of $\gamma$. This feature makes it possible for the electron to experience the increased energy gain.

\begin{figure}[t]
    \begin{center}
    \includegraphics[width=1\columnwidth,clip]{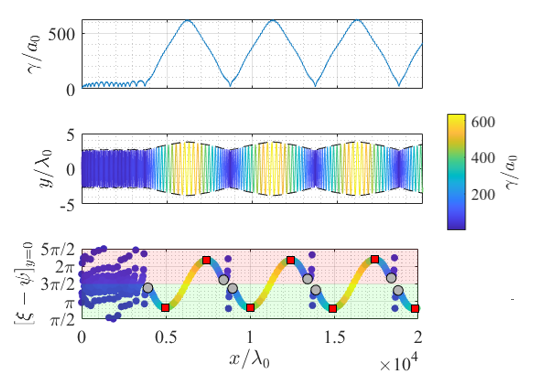}
    \caption{\label{fig:alphaConstPhase} Electron dynamics for the case with $S = 9.5$ and $\delta u = \delta u_{\max}$ ($\alpha = 1.4$ and $a_0 = 8$). (a) $\gamma$ as a function of the longitudinal position $x$. (b) Electron trajectory $y(x)$, with the color indicating $\gamma/a_0$. The dashed lines show $\pm y_*$, where $y_*$ is the amplitude of the betatron oscillations given by \cref{eq:y_*}. (c) Phase offset between the laser phase and the betatron phase, $[\xi - \psi]_{y=0}$, evaluated on the axis. The gray and red markers indicate locations where $\langle \omega' \rangle = \omega_{\beta}$.}
    \end{center}
\end{figure}

The energy-gain band has a finite width because the detuning is not a binary process. In fact, what limits the energy gain is the gradual accumulation of a phase offset between the laser field and the betatron motion. If the frequency mismatch is relatively small, then this accumulation can be slow. On the other hand, a significant mismatch speeds up the accumulation.

\begin{figure}[t]
    \begin{center}
    \includegraphics[width=1\columnwidth,clip]{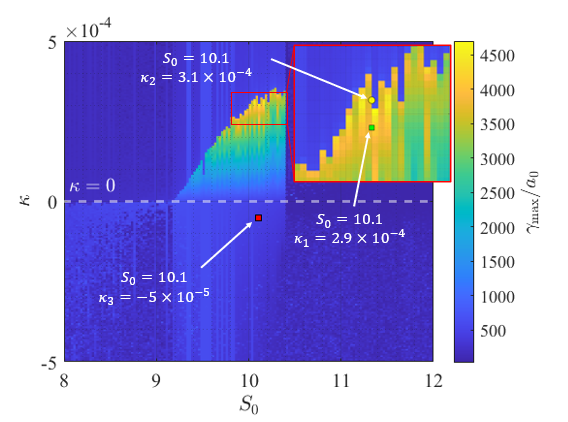}
    \caption{\label{fig:kappaLinear} Maximum $\gamma$ attained by an electron as a function of the initial value of $S$, denoted as $S_0$, and the normalized gradient of $\alpha$, denoted as $\kappa$. We use the same value $\delta u = \delta u_{\max}$ and the same initial conditions for each $S_0$ as those used in the scan of \cref{fig:duMax}, including the corresponding initial laser phase. The markers indicate \emph{Cases 1--3} discussed in \cref{sec-linear}.}
    \end{center}
\end{figure}

To make this argument quantitative, we introduce the betatron phase $\psi$. As shown in Ref.~[\onlinecite{yeh.pop.2025}], the transverse electron motion is confined to $|y| \leq y_*$, where
\begin{equation} \label{eq:y_*}
    y_* = \left[ \frac{S + \gamma \delta u}{u \alpha} \right]^{1/2} \lambda_0 
\end{equation}
is the local amplitude of betatron oscillations. The betatron phase $\psi$ is then defined through
\begin{equation}
    y = y_* \sin(\psi).
\end{equation}
A convenient diagnostic is the phase offset between the laser phase $\xi$ and the betatron phase $\psi$ evaluated on the axis of the current filament, $[\xi - \psi]_{y = 0}$. This quantity is easy to calculate, because $\psi = 0$ on the axis. It also provides a direct way to evaluate the frequency ratio during numerical integration of the equations of motion, as described in Ref.~[\onlinecite{yeh.pop.2025}].

The evolution of $\gamma$, $y$, and the phase offset $[\xi - \psi]_{y = 0}$ for an electron with $S = 9.5$, plotted as functions of the longitudinal position $x$, is shown in \cref{fig:alphaConstPhase}. Note that the very same electron trajectory was used to evaluate the frequency ratio shown in \cref{fig:alphaConstFreq}. As the electron gains energy, the phase offset remains within the interval between $\pi/2$ and $3\pi/2$, which has been shown to correspond to a favorable phase relation that produces net energy gain over a betatron oscillation~\cite{yeh.pop.2025,arefiev.pop.2024}. The key point is that detuning and frequency mismatch, shown in \cref{fig:alphaConstFreq}, lead to a gradual evolution of the phase offset rather than an abrupt loss of energy gain. 

The nonmonotonic dependence of the frequency ratio shown in \cref{fig:alphaConstFreq}, which results from the superluminal phase velocity, delays the phase offset from exiting the favorable range. To illustrate this, the purple and red markers in \cref{fig:alphaConstPhase}(c) indicate locations where the betatron resonance condition is satisfied exactly. Consider the change in the phase offset associated with the first peak in $\gamma$. Immediately after the first resonance crossing (purple circle), the frequency ratio drops below unity and the phase offset begins to decrease. However, before the offset reaches $\pi/2$, the frequency ratio undergoes a second resonance crossing (red square) and then exceeds 1. As a result, the phase offset begins to increase, which keeps it within the favorable range. As the frequency ratio continues to grow, the phase offset eventually exits the favorable range and the energy gain terminates.

We can now revisit the energy-gain band in \cref{fig:duMax} and provide a qualitative explanation for this feature that will be useful as we move forward. For values of $S$ on the left side of the band, the electron passes through the resonance only once. In this case, the phase offset begins to decrease as the frequency ratio drops below unity, but the reversal of this trend does not occur before the phase offset exits the favorable range and the electron starts losing energy. For values of $S$ on the right side of the band, the electron never reaches the resonance condition at all: the frequency ratio remains above unity, and the energy gain stops before the resonance can even be established. Only for intermediate values of $S$ does the electron reverse the evolution of the phase offset after crossing the resonance for the second time while remaining within the favorable range, which leads to sustained energy gain and the formation of the energy-gain band.

\begin{figure}[t]
    \begin{center}
    \includegraphics[width=1\columnwidth,clip]{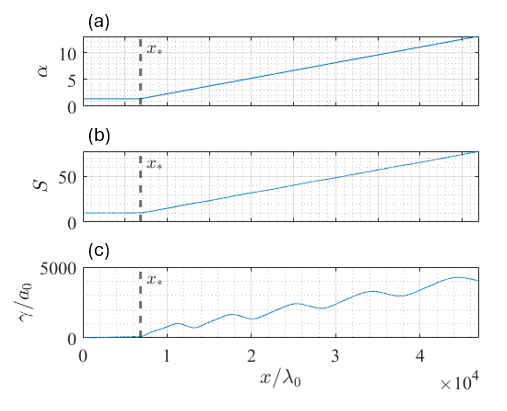}
    \caption{\label{fig:HighELinear} Evolution of $\alpha$, $S$, and $\gamma$ along $x$ for \emph{Case~1} from \cref{fig:kappaLinear}, with $\kappa=2.90 \times 10^{-4}$ and $S_0=10.1$.}
    \end{center}
\end{figure}

In summary, a slightly superluminal laser phase velocity can delay the onset of significant frequency detuning and thereby enable substantially higher electron energies than in the luminal case. This delay allows the phase offset to remain within the favorable range for an extended interval, resulting in sustained energy gain over many betatron oscillations. However, even with superluminosity, the frequencies ultimately detune as the electron energy increases. This causes the phase offset to grow until it exits the favorable range, at which point the energy gain terminates and is followed by energy loss. This unavoidable limitation of the longitudinally uniform case motivates the exploration of cases with a longitudinally varying current parameter $\alpha$.


\section{Electron dynamics in a plasma magnetic field with a linear field-strength ramp}
\label{sec-linear}

One defining feature of the electron dynamics at constant $\alpha$ is that the dependence of the frequency ratio on $\gamma$ is fixed. By changing $\alpha$ along the electron trajectory, we can alter the evolution of the frequency ratio, so that it is no longer uniquely prescribed by the electron's $\gamma$. In this section, we show that this gives us the ability to regulate the evolution of the offset between the laser phase and the betatron phase, thereby substantially prolonging the energy gain and increasing the $\gamma$ attained by the electron. We start by presenting a scan that shows the change in the attained $\gamma$ as a function of the longitudinal gradient of $\alpha$ and then focus on three representative cases from this scan. 

The scan in \cref{fig:kappaLinear} shows how the rate at which $\alpha$ changes along $x$ affects the maximum $\gamma$ attained by the electron. To make a clear comparison with the electron dynamics in a uniform $\alpha$, we used the scan from \cref{fig:duMax} as a baseline. The scan in \cref{fig:duMax} was performed for an optimal $\delta u = \delta u_{\max}$ by varying $S$ and the initial laser phase that was chosen for each $S$ to maximize the value of $\gamma$ attained by the electron. In the scan shown in \cref{fig:kappaLinear}, we keep all initial values exactly the same as in \cref{fig:duMax}, including the initial value of $\alpha$ that we now refer to as $\alpha_0$. The integration time is also the same. The only change that we make is to introduce a linear dependence of $\alpha$ on $x$. This dependence is represented by $\kappa$, which is the normalized gradient of $\alpha$. Furthermore, even though $S$ is generally no longer conserved, we use the initial value of $S$, denoted as $S_0$, as a scan parameter. This way the line-out at $\kappa = 0$ in \cref{fig:kappaLinear} is the scan from \cref{fig:duMax}.

\begin{figure}[b]
    \begin{center}
    \includegraphics[width=1\columnwidth,clip]{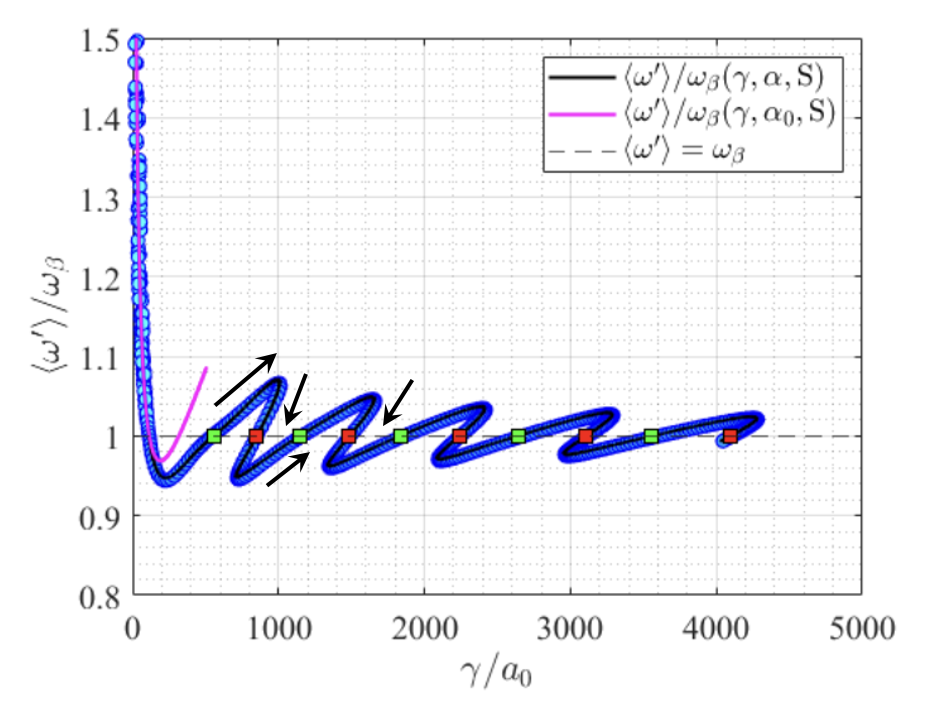}
    \caption{\label{fig:HighELinearFreq} The frequency ratio for \emph{Case~1}: the blue filled circles show values obtained during numerical integration of the equations of motion and the black solid curve is the analytical prediction from \cref{eq:freq_ratio}. The successive resonance crossings after $\alpha$ begins to increase are labeled with alternating green and red markers. The magenta curve is the analytical prediction from \cref{eq:freq_ratio} for $\alpha = \alpha_0$.}
    \end{center}
\end{figure}

We use a piecewise-linear profile for $\alpha(x)$, which allows us to control the point $x_*$ at which $\alpha$ begins to change:
\begin{equation} \label{eq-linProfile}
\alpha(x) =
    \begin{cases} 
          \alpha_0, & x \leq x_*, \\
          \kappa (x - x_*) / \lambda_0 + \alpha_0, & x > x_*,
    \end{cases}
\end{equation}
where $\kappa$ is the normalized gradient. As illustrated in \cref{fig:alphaConstPhase}(a), the energy gain does not necessarily begin immediately in the longitudinally uniform case. We therefore introduce the variation of $\alpha$ only after the onset of energy gain, at $x=x_*$, so that the effect of the longitudinal ramp is tied directly to the subsequent energy-increase stage. The procedure used to set $x_*$ depends on the initial value of $S$. For values of $S$ inside the energy-gain band shown in \cref{fig:duMax}, as well as for values of $S$ to the left of the band, we take $x_*$ as the first longitudinal location where the betatron resonance condition is satisfied. For values of $S$ to the right of the band, the electron never reaches resonance. In this case, we analyze the corresponding constant-$\alpha$ trajectory, identify the peak in $\gamma(x)$ that contains $\gamma_{\max}$, and set $x_*$ at the beginning of that peak.

\begin{figure}[t]
    \begin{center}
    \includegraphics[width=1\columnwidth,clip]{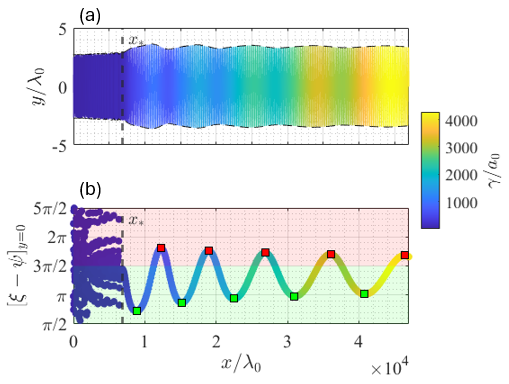}
    \caption{\label{fig:HighELinearPhase} The electron trajectory (a) and phase offset between the laser phase and the betatron phase (b) for \emph{Case~1}. In panel (a), the dashed curves show $\pm y_*$, where $y_*$ is the betatron amplitude given by \cref{eq:y_*}. In panel (b), the alternating green and red markers show successive resonance crossings with $\langle \omega' \rangle/\omega_{\beta}=1$ after $\alpha$ begins to increase (the same crossings as in \cref{fig:HighELinearFreq}). The green shaded region indicates the favorable range and the red shaded region indicates the unfavorable range of the phase offset.}
    \end{center}
\end{figure}

The key finding of the scan is that increasing $\alpha$ ($\kappa > 0$) can substantially increase $\gamma_{\max}$, in some cases by several fold compared to the longitudinally uniform case. In contrast, decreasing $\alpha$ ($\kappa<0$) produces no comparable increase and often reduces $\gamma_{\max}$. For sufficiently large positive values of $\kappa$, $\gamma_{\max}$ eventually decreases, indicating the existence of an optimal range of gradients. Finally, the enhancement provided by $\kappa>0$ is confined to the same interval of $S_0$ that corresponds to the energy-gain band in the constant-$\alpha$ scan shown in \cref{fig:duMax}. The longitudinal variation of $\alpha$ is therefore particularly effective when altering the trajectories that already result in enhanced energy gain in the longitudinally uniform case.

\begin{figure}[t]
    \begin{center}
    \includegraphics[width=1\columnwidth,clip]{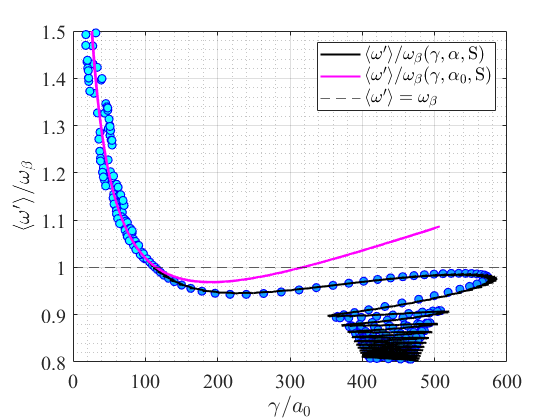}
    \caption{\label{fig:OvershootFreq} The frequency ratio for \emph{Case~2}: the blue filled circles show values obtained during numerical integration of the equations of motion and the black solid curve is the analytical prediction from \cref{eq:freq_ratio}. The magenta curve is the analytical prediction from \cref{eq:freq_ratio} for $\alpha = \alpha_0$.}
    \end{center}
\end{figure}

We next examine the electron dynamics to pinpoint how the longitudinal variation of $\alpha$ impacts the energy-gain process. To do that, we selected three representative cases, marked in \cref{fig:kappaLinear}, that have the same $S_0 = 10.1$ and differ only by their value of $\kappa$. The first case (\emph{Case 1}) has $\gamma_{\max}$ that is close to the highest value attained in our scan, with $\kappa = 2.90 \times 10^{-4}$. The second case (\emph{Case 2}) is just outside the optimal range of $\kappa$, with $\kappa = 3.10 \times 10^{-4}$, and does not exhibit a significant increase of $\gamma_{\max}$. Finally, the third case (\emph{Case 3}) corresponds to a decreasing current parameter, with $\kappa = -5 \times 10^{-5}$.

\begin{figure}[t]
    \begin{center}
    \includegraphics[width=1\columnwidth,clip]{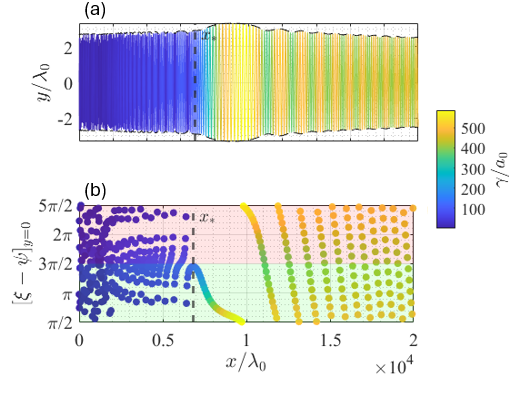}
    \caption{\label{fig:OvershootPhase} The electron trajectory (a) and phase offset between the laser phase and the betatron phase (b) for \emph{Case~2}. In panel (a), the dashed curves show $\pm y_*$, where $y_*$ is the betatron amplitude given by \cref{eq:y_*}. In panel (b), the green shaded region indicates the favorable range and the red shaded region indicates the unfavorable range of the phase offset.}
    \end{center}
\end{figure}

\Cref{fig:HighELinear} shows $\gamma(x)$ and $S(x)$ for \emph{Case 1}. For reference, this figure also shows $\alpha(x)$. The most striking feature is the qualitative change in $\gamma(x)$ compared to the case with constant $\alpha$. As seen in \cref{fig:alphaConstPhase}(a), $\gamma(x)$ exhibits a sequence of comparable peaks separated by deep drops in $\gamma$ when $\alpha$ is constant. The increasing $\alpha$ causes the peaks to grow progressively, with $\gamma$ not returning to low values. This growth without saturation indicates that the $\gamma_{\max}$ shown in \cref{fig:kappaLinear} may not even be the highest attainable $\gamma$ in this configuration.

Using the electron dynamics, we computed the frequency ratio at each position along $x$. \Cref{fig:HighELinearFreq} shows this ratio (blue markers) as a function of the electron's $\gamma$. In contrast to the case with constant $\alpha$ that is shown in magenta, this ratio is no longer uniquely prescribed by the electron's $\gamma$. The magenta curve is generated from the analytical expression given by \cref{eq:freq_ratio} using $S=S_0$ and $\alpha=\alpha_0$. The same analytical expression can also predict the frequency ratio for a changing $\alpha$, provided we use the instantaneous values of $\alpha$, $S$, and $\gamma$. The result, shown with the black curve, agrees well with the values obtained directly from the electron dynamics. The arrows indicate how the frequency ratio evolves along the electron trajectory and show that, as $\alpha$ increases along the trajectory, the frequency ratio is pushed downward. This raises two questions: what causes this trend and how is it linked to the evolution of $\gamma$? We address these questions in sequence by first focusing on the origin of the trend.

The behavior of the frequency ratio can be understood using \cref{eq:freq_ratio}. At sufficiently large $\gamma$, the second term inside the brackets dominates and the frequency ratio can be approximated as
\begin{equation}
    \frac{\langle \omega' \rangle}{\omega_{\beta}} \approx \frac{3 \pi}{\sqrt{2 \alpha}} \gamma^{1/2} \delta u ,
    \label{eq:freq_ratio-approx}
\end{equation}
so that $\langle \omega' \rangle/ \omega_{\beta} \propto \sqrt{\gamma / \alpha}$. At constant $\alpha$, the electron then retraces the same curve in the $(\gamma,\langle \omega' \rangle/\omega_{\beta})$ plane as it gains and loses energy. If $\alpha$ changes along the electron trajectory, as is the case in our example, the frequency ratio exhibits hysteresis, in the sense that its value at a given $\gamma$ becomes dependent on the electron's prior motion. Specifically, since $\alpha$ increases during the electron motion, the electron returns to a given $\gamma$ with a lower frequency ratio, which shifts the curve in the $(\gamma,\langle \omega' \rangle/\omega_{\beta})$ plane downward.

\begin{figure}[t]
    \begin{center}
    \includegraphics[width=1\columnwidth,clip]{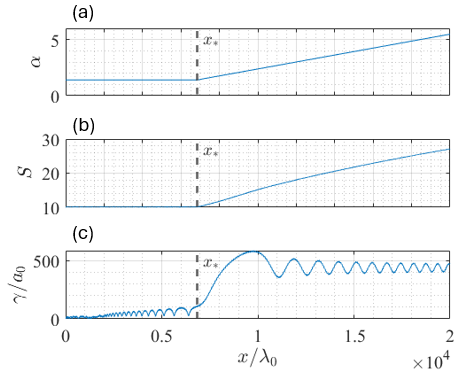}
    \caption{\label{fig:overshootPlots} Evolution of $\alpha$, $S$, and $\gamma$ along $x$ for \emph{Case~2} from \cref{fig:kappaLinear}, with $\kappa=3.10 \times 10^{-4}$ and $S_0=10.1$.}
    \end{center}
\end{figure}

To understand how the hysteresis influences the energy-gain process, we must consider its impact on the evolution of the phase offset. The value of the phase offset directly determines whether the electron gains or loses energy. In the range between $\pi/2$ and $3\pi/2$, the electron gains energy, so we refer to this range as favorable. In the range between $3 \pi/2$ and $5\pi/2$, the electron loses energy, so we refer to this range as unfavorable. The phase offset as a function of $x$ for the considered example is shown in \cref{fig:HighELinearPhase}(b). We use it in our subsequent analysis in conjunction with the frequency ratio plot in \cref{fig:HighELinearFreq}. To make the correlation between \cref{fig:HighELinearFreq} and \cref{fig:HighELinearPhase} easier to see, we label successive resonance crossings after $\alpha$ begins to increase using alternating green and red markers.

We first consider the electron dynamics between the first green and red markers. Between these markers, the corresponding segment of the curve in \cref{fig:HighELinearFreq} has $\langle\omega'\rangle/\omega_{\beta}>1$, so the phase offset (see \cref{fig:HighELinearPhase}) increases as the electron first gains and then loses energy. The electron's $\gamma$ starts to decrease once the phase offset enters the unfavorable range. This decrease causes the frequency ratio to drop. However, since $\alpha$ is increasing at the same time, the frequency ratio in \cref{fig:HighELinearFreq} is pushed down more rapidly and reaches $\langle\omega'\rangle/\omega_{\beta}=1$ at a higher $\gamma$ than at the first green marker. This return to unity corresponds to the first red marker.

The electron reaches the first red marker with the phase offset in the unfavorable range, so $\gamma$ continues to drop. As a result, the frequency ratio continues to decrease and becomes less than unity. At this point, the phase offset reverses its direction and begins to decrease toward the favorable range. Because the frequency ratio is pushed downward rapidly by the increasing $\alpha$, the electron reaches the favorable range while the frequency ratio is still below unity. Once the phase offset re-enters the favorable range, $\gamma$ starts to increase, which produces the kink in \cref{fig:HighELinearFreq}. In contrast, in the case of uniform $\alpha$, the frequency ratio returns to unity while the phase offset is still in the unfavorable range. The phase offset then reverses direction before it can reach the favorable range, leading to a continued decrease in $\gamma$ and, ultimately, to deep energy losses.

\begin{figure}[t]
    \begin{center}
    \includegraphics[width=1\columnwidth,clip]{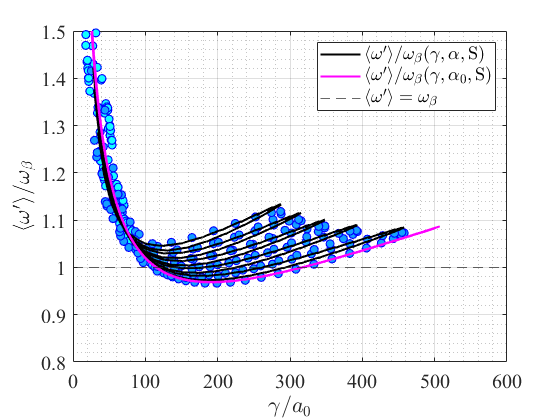}
    \caption{\label{fig:NegativeFreq} The frequency ratio for \emph{Case~3}: the blue filled circles show values obtained during numerical integration of the equations of motion and the black solid curve is the analytical prediction from \cref{eq:freq_ratio}. The magenta curve is the analytical prediction from \cref{eq:freq_ratio} for $\alpha = \alpha_0$. }
    \end{center}
\end{figure}

After the kink, the frequency ratio remains below unity, so the phase offset continues to decrease and moves deeper into the favorable range while $\gamma$ increases. As $\gamma$ grows, the scaling in \cref{eq:freq_ratio-approx} implies a tendency for $\langle\omega'\rangle/\omega_{\beta}$ to increase. However, $\alpha$ is increasing at the same time, which counteracts this tendency and delays the return of the frequency ratio to unity. As a result, the electron does not retrace its path in \cref{fig:HighELinearFreq} back toward the first red marker. Instead, the frequency ratio returns to unity at a $\gamma$ higher than at the first red marker. This new resonance crossing is the second green marker.

The described green-red-green marker sequence repeats as $\alpha$ continues to increase. The frequency ratio $\langle \omega' \rangle / \omega_{\beta}$ continues to cycle through successive local maxima and minima, with both shifting to progressively higher values of $\gamma$. As a result, the dependence of $\langle \omega' \rangle / \omega_{\beta}$ on $\gamma$ develops the zig-zag--like shape seen in \cref{fig:HighELinearFreq}. The progressive shift of both the local minima and maxima of the frequency ratio to higher $\gamma$ manifests as successive peaks in $\gamma(x)$ that reach progressively higher values in \cref{fig:HighELinear}(c), without the pronounced drop in $\gamma$ between peaks that is characteristic of the longitudinally uniform case.

In the considered example, the increase of $\alpha$ is beneficial. It counteracts the increase of the frequency ratio caused by increasing $\gamma$ and amplifies the reduction of the frequency ratio caused by decreasing $\gamma$. However, the increase of $\alpha$ becomes harmful if it is too rapid, as manifested by the threshold in \cref{fig:kappaLinear} at high positive $\kappa$.

The mechanism responsible for the threshold is illustrated by the next example (yellow marker in the inset of \cref{fig:kappaLinear}), which we earlier labeled as \emph{Case 2}. Here, $\kappa = 3.10\times 10^{-4}$ is higher than in \emph{Case 1} we have just discussed. The corresponding frequency ratio as a function of $\gamma$ is shown in \cref{fig:OvershootFreq}, together with the curve for constant $\alpha = \alpha_0$. The linear increase of $\alpha$ starts once $\langle \omega' \rangle / \omega_{\beta}$ drops below unity for the first time while the electron is gaining energy. This means that the phase offset is in the favorable range and is decreasing (see \cref{fig:OvershootPhase}). The subsequent dynamics strongly depends on the competition between the frequency ratio returning toward unity and the phase offset reaching the edge of the favorable range at $\pi/2$. In \emph{Case 1}, $\langle \omega' \rangle / \omega_{\beta}$ returns to unity first, so the direction of the phase offset reverses and the electron can keep gaining energy. In the current example, the phase offset reaches $\pi/2$ first. As a result, the phase offset enters the unfavorable range and $\gamma$ starts to decrease, which prevents the frequency ratio from reaching the second resonance crossing. The failure to reach the second resonance crossing is what produces the threshold behavior in \cref{fig:kappaLinear}.

\begin{figure}[t]
    \begin{center}
    \includegraphics[width=1\columnwidth,clip]{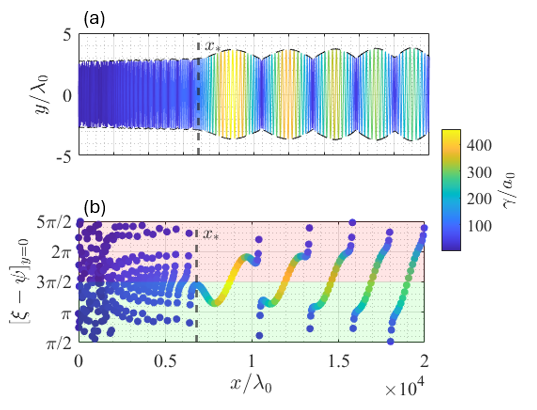}
    \caption{\label{fig:NegativePhase} The electron trajectory (a) and phase offset between the laser phase and the betatron phase (b) for \emph{Case~3}. In panel (a), the dashed curves show $\pm y_*$, where $y_*$ is the betatron amplitude given by \cref{eq:y_*}. In panel (b), the green shaded region indicates the favorable range and the red shaded region indicates the unfavorable range of the phase offset.}
    \end{center}
\end{figure}

After $\gamma$ starts to decrease, the continued increase of $\alpha$ prevents deep energy losses and instead enables the electron to retain a significant portion of its energy. The key factor here is the downward push of $\langle \omega' \rangle / \omega_{\beta}$ further away from unity. The push causes the phase offset to decrease more and more rapidly, as seen in \cref{fig:OvershootPhase}(b). It repeatedly traverses the unfavorable and favorable ranges, which makes it impossible for the electron to lose a large fraction of its energy. Instead, its $\gamma$ fluctuates, as seen in \cref{fig:overshootPlots}(c), with the amplitude of the fluctuations decreasing, while the characteristic value remains around a moderate level. This behavior contrasts with the case of constant $\alpha$, where the electron retraces its trajectory in the $(\gamma,\langle \omega' \rangle / \omega_{\beta})$ plane, which leads to deep energy losses.

We now switch our attention to regimes where $\kappa<0$, so $\alpha$ decreases with $x$. In these regimes, lowering $\alpha$ pushes the frequency ratio up. One might expect this to be beneficial in cases where, at constant $\alpha$, the electron fails to reach the second resonance crossing while the phase offset remains in the favorable range. These are the cases that correspond to values of $S_0$ in \cref{fig:kappaLinear} to the left of the energy-gain band. Indeed, pushing the frequency-ratio curve upward shifts the second resonance crossing predicted by \cref{eq:freq_ratio} to lower $\gamma$. This shift can help the electron reach the second resonance crossing. However, the tradeoff is that the crossing occurs at lower $\gamma$, which reduces its ability to significantly increase $\gamma_{\max}$. The scan in \cref{fig:kappaLinear} shows that indeed no significant increase of $\gamma_{\max}$ occurs to the left of the energy-gain band at $\kappa<0$.

\begin{figure}[b]
    \begin{center}
    \includegraphics[width=1\columnwidth,clip]{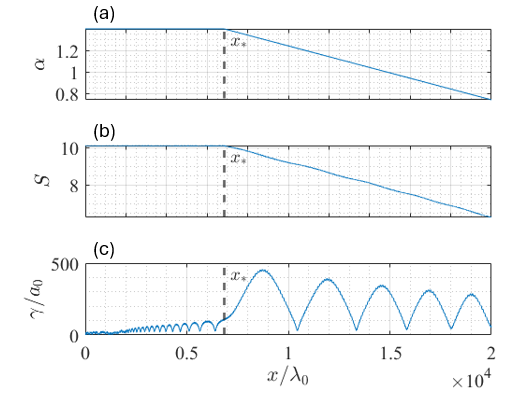}
    \caption{\label{fig:negativePlots} Evolution of $\alpha$, $S$, and $\gamma$ along $x$ for \emph{Case~3} from \cref{fig:kappaLinear}, with $\kappa = -5 \times 10^{-5}$ and $S_0=10.1$.}
    \end{center}
\end{figure}

To examine the impact of lowering $\alpha$ on electron dynamics within the energy-gain band of the uniform-$\alpha$ case, we turn our attention to \emph{Case 3} that is marked in red in \cref{fig:kappaLinear}. The corresponding frequency ratio as a function of $\gamma$ is shown in \cref{fig:NegativeFreq}. The linear decrease of $\alpha$ starts after the first resonance crossing, once $\langle \omega' \rangle / \omega_{\beta}$ drops below unity for the first time while the electron is gaining energy. This means that the phase offset is in the favorable range and is decreasing (see \cref{fig:NegativePhase}). As discussed above, the upward push of $\langle \omega' \rangle / \omega_{\beta}$ shifts the second resonance crossing to lower $\gamma$, so the phase offset reverses direction at somewhat lower $\gamma$. The most profound impact, however, occurs after the second resonance crossing, when $\langle \omega' \rangle / \omega_{\beta}$ rises above unity again. Once the increasing phase offset enters the unfavorable range, it is important to reverse its direction as soon as possible, before $\gamma$ drops significantly. In \emph{Case 1}, the downward push of $\langle \omega' \rangle / \omega_{\beta}$ caused by increasing $\alpha$ allowed it to drop below unity quickly once $\gamma$ started to decrease, so the phase offset could reverse direction again and exit the unfavorable range. In contrast, this exit is delayed in the current example because of the upward push of $\langle \omega' \rangle / \omega_{\beta}$. In fact, it never occurs before the electron loses most of its energy, as seen in \cref{fig:NegativePhase}(b).

After the energy loss, the electron re-enters the energy-gain stage and the described scenario repeats. As a result, $\gamma(x)$ in \cref{fig:negativePlots}(c) exhibits distinct peaks with deep reductions of $\gamma$ in between. Each new energy-gain cycle begins with lower $\alpha$ and lower $S$, as seen in \cref{fig:negativePlots}. These changes shift the $\langle \omega' \rangle / \omega_{\beta}$ versus $\gamma$ curve upward. As $\alpha$ and $S$ continue to decrease, this upward shift eventually becomes strong enough that the electron can no longer enter the betatron resonance. In particular, the minimum of $\langle \omega' \rangle / \omega_{\beta}$ rises above unity, so the condition $\langle \omega' \rangle / \omega_{\beta}=1$ is no longer satisfied. For the constant-$\alpha$ case, Ref.~[\onlinecite{yeh.pop.2025}] gives $\min(\langle \omega' \rangle / \omega_{\beta}) \propto \sqrt{S/\alpha}$. In the present case, $\alpha$ decreases faster than $S$, so this minimum increases and eventually exceeds unity.

Overall, the results of this section identify the longitudinal increase of $\alpha$ as a control knob for regulating the phase offset and, therefore, the energy-gain process. The control is achieved through a previously unexplored hysteresis of $\langle\omega'\rangle/\omega_{\beta}$. Specifically, the variation of $\alpha$ imprints the electron's prior motion onto the dependence of $\langle\omega'\rangle/\omega_{\beta}$ on $\gamma$.

\begin{figure}[t]
    \begin{center}
    \includegraphics[width=1\columnwidth,clip]{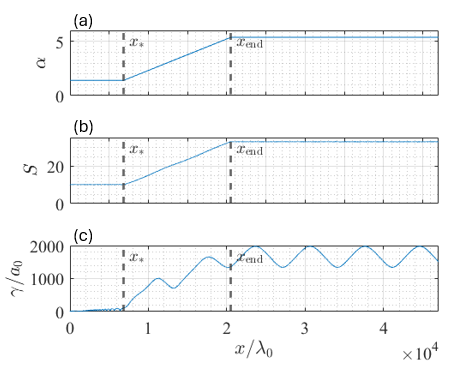}
    \caption{\label{fig:rampPlots} Evolution of $S$ and $\gamma$ along $x$ for a modified $\alpha$ profile that matches $\alpha(x)$ for \emph{Case~1} from \cref{fig:HighELinear} at $x \leq x_{\mathrm{end}}$ and then becomes constant at $x > x_{\mathrm{end}}$.}
    \end{center}
\end{figure}


\section{Breaking reversibility in direct laser acceleration} \label{sec: regulation}

In \cref{sec-linear}, we uncovered the potential of regulating the phase offset using the hysteresis dependence of $\langle\omega'\rangle/\omega_{\beta}$ on $\gamma$. This mechanism gives us the ability to qualitatively change the electron dynamics. Specifically, it allows us to break what is typically perceived as a hallmark of direct laser acceleration: the largely reversible energy exchange between the electron and the laser field. In this section, we highlight two important outcomes of this break in reversibility: energy retention and steady energy gain without intermittent losses.

\begin{figure}[t]
    \begin{center}
    \includegraphics[width=1\columnwidth,clip]{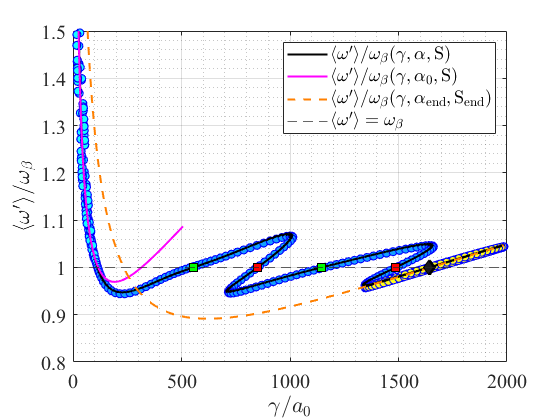}
    \caption{\label{fig:RampFreq} The frequency ratio for the modified $\alpha$ profile from \cref{fig:rampPlots}(a). Circles show values obtained during numerical integration of the equations of motion, with filled blue circles for $x \leq x_{\mathrm{end}}$ and open circles for $x > x_{\mathrm{end}}$. Resonance crossings for $x_*<x\leq x_{\mathrm{end}}$ are labeled with alternating green and red markers and the crossing for $x>x_{\mathrm{end}}$ is shown with a black diamond. The black solid curve is the analytical prediction from \cref{eq:freq_ratio}. The magenta curve is the analytical prediction for $\alpha=\alpha_0$ and $S=S_0$ and the orange dashed curve is the analytical prediction for $\alpha=\alpha_{\mathrm{end}}$ and $S=S_{\mathrm{end}}$.}
    \end{center}
\end{figure}

The reversibility during direct laser acceleration is illustrated in \cref{fig:alphaConstPhase}(a), which shows $\gamma$ as a function of $x$ in the case of a constant $\alpha$. The profile of the electron's $\gamma$ consists of distinct peaks, with a significant increase in $\gamma$ always followed by a significant decrease. This means that, if an electron that is gaining energy is given enough time, it will eventually start losing energy.  

In \cref{sec-linear}, we considered a setup where $\alpha$ increases linearly with $x$. In the example labeled as \emph{Case~1}, the normalized gradient $\kappa = 2.90 \times 10^{-4}$ led to a significant increase of the electron $\gamma$ compared to the uniform-$\alpha$ case. The evolution of $\alpha$ and the corresponding evolution of $\gamma$ are shown in \cref{fig:HighELinear}. Over the simulated interval, the envelope of the $\gamma$ peaks showed no clear saturation in this regime.

To test whether the electron can retain the energy gained during the increase of $\alpha$, we use a modified $\alpha(x)$ while keeping all other parameters and initial conditions the same as in \emph{Case~1}. As shown in \cref{fig:rampPlots}(a), the linear increase stops at $x_{\mathrm{end}}$, so $\alpha$ and thus also $S$ become constant again at $x>x_{\mathrm{end}}$, with $\alpha=\alpha_{\mathrm{end}}$ and $S=S_{\mathrm{end}}$. The corresponding $\gamma(x)$ is shown in \cref{fig:rampPlots}(c). At $x>x_{\mathrm{end}}$, the peaks of $\gamma$ no longer increase and $\gamma$ instead oscillates around a new elevated level. This means that the electron is able to retain most of the energy it acquired during the linear increase of $\alpha$.

\begin{figure}[t]
    \begin{center}
    \includegraphics[width=1\columnwidth,clip]{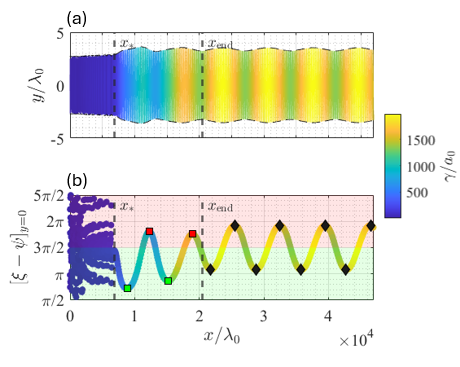}
    \caption{\label{fig:RampPhase} The electron trajectory (a) and phase offset between the laser phase and the betatron phase (b) for the modified $\alpha$ profile from \cref{fig:rampPlots}(a). In panel (a), the dashed curves show $\pm y_*$, where $y_*$ is the betatron amplitude given by \cref{eq:y_*}. In panel (b), the green shaded region indicates the favorable range and the red shaded region indicates the unfavorable range of the phase offset. The square and diamond markers are the locations where $\langle \omega' \rangle/\omega_{\beta}=1$. The labeling matches that in \cref{fig:RampFreq}.}
    \end{center}
\end{figure}

To understand why the electron retains its energy, we examine the dependence of $\langle \omega' \rangle / \omega_{\beta}$ on $\gamma$ at $x > x_{\mathrm{end}}$. The orange dashed curve in \cref{fig:RampFreq} shows the analytical prediction given by \cref{eq:freq_ratio} for $\alpha=\alpha_{\mathrm{end}}$ and $S=S_{\mathrm{end}}$. The ratio calculated using the electron dynamics (open markers) follows this curve at $x > x_{\mathrm{end}}$. The electron moves to the right along the curve when the phase offset is in the favorable range, whereas it moves to the left when the offset is in the unfavorable range. The dashed curve lies below unity between $\gamma \approx 1650$ and $\gamma \approx 250$, so $\langle \omega' \rangle / \omega_{\beta}<1$ throughout the interval over which $\gamma$ would have to decrease substantially. Therefore, a significant reduction of $\gamma$ would require the phase offset to keep decreasing while remaining in the unfavorable range. This proves to be impossible for the electron, as shown in \cref{fig:RampPhase}(b). The phase offset decreases into the favorable range before $\gamma$ can drop appreciably, which interrupts the energy-loss stage and restores energy gain.

The lack of reversibility that leads to energy retention can also be viewed in terms of the accessibility of the attained high-$\gamma$ state when starting from low $\gamma$ at constant $\alpha$. The original curve ($\alpha=\alpha_0$ and $S=S_0$), shown in magenta, has a much smaller separation in $\gamma$ between the two resonance crossings. This is why the electron is able to reach the second resonance crossing while gaining energy. In the case of the orange dashed curve, the separation is much wider. Starting from low $\gamma$, the electron is unable to follow this curve and reach the second resonance crossing. In fact, the electron is not even able to reach the first resonance crossing. We verified this by directly simulating the electron dynamics with $\alpha=\alpha_{end}$ and $S=S_{end}$ and optimizing the initial laser phase. The key conclusion is that the hysteresis during the increase of $\alpha$ transfers the electron onto a high-$\gamma$ trajectory that cannot be reached by evolving at constant $\alpha$ starting from low $\gamma$.

\begin{figure}[t]
    \begin{center}
    \includegraphics[width=1\columnwidth,clip]{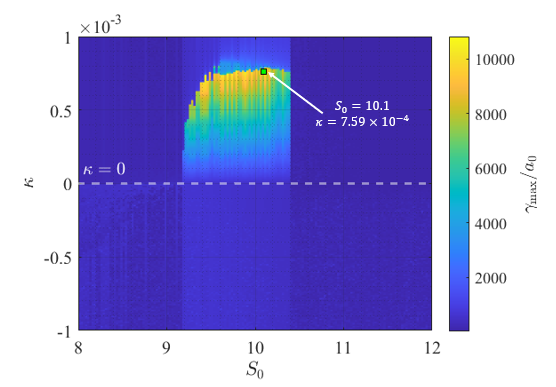}
    \caption{\label{fig:kappaLinearXStarScan} A parameter scan similar to that in \cref{fig:kappaLinear} except that $x_*$ is chosen as the second resonance crossing for values of $S_0$ within the energy-gain band. The green square marker shows the parameters for the example in \cref{fig:xStarPlots,fig:xStarFreq,fig:xStarPhase}.}
    \end{center}
\end{figure}

\begin{figure}[b]
    \begin{center}
    \includegraphics[width=1\columnwidth,clip]{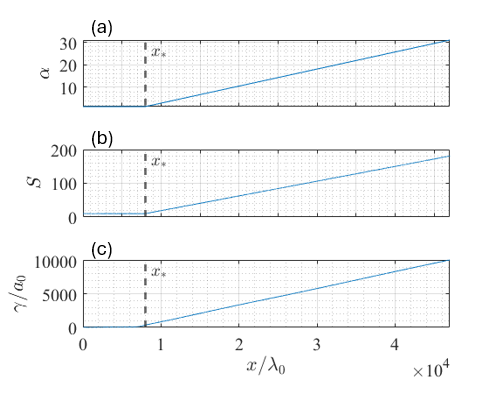}
    \caption{\label{fig:xStarPlots} Evolution of $\alpha$, $S$, and $\gamma$ along $x$ for an example from the scan in \cref{fig:kappaLinearXStarScan} with  $\kappa=7.59 \times10^{-4}$ and $S_0=10.1$.}
    \end{center}
\end{figure}

The hysteresis produced by a changing $\alpha$ breaks the typical symmetry between energy gain and energy loss during DLA. In the example discussed above, this asymmetry allows the electron to access a high-energy state that is not reachable when $\alpha$ is uniform. This naturally raises the question of whether the intermittent energy-loss stages can be eliminated. Up to this point, we have focused on the role of the normalized gradient $\kappa$, but we also have the freedom to choose where the increase of $\alpha$ is turned on by selecting $x_*$. The next example leverages this control to obtain a regime in which the energy gain becomes continuous.

\Cref{fig:kappaLinearXStarScan} shows a scan where we changed the onset of the longitudinal variation of $\alpha$ compared to the scan in \cref{fig:kappaLinear}. The change only applies to values of $S_0$ within the energy-gain band. Recall that the distinctive feature of the energy-gain band is that $\langle\omega'\rangle/\omega_{\beta}$ goes through resonance twice as $\gamma$ increases during the energy gain stage. In this new scan, we take $x_*$ to be the location of the second rather that first resonance crossing. The change of $x_*$ has a strong impact on the scan: the $\kappa$ threshold and the highest $\gamma_{\max}$ increase by a factor of two. The threshold increase is particularly interesting, because now we leverage higher values of $\kappa$ that were previously considered counterproductive to achieve even higher values of $\gamma$.

\begin{figure}[t]
    \begin{center}
    \includegraphics[width=1\columnwidth,clip]{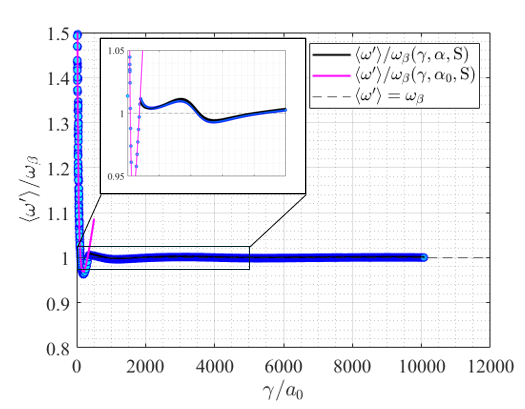}
    \caption{\label{fig:xStarFreq} The frequency ratio for an example from the scan in \cref{fig:kappaLinearXStarScan} with $\kappa=7.59 \times10^{-4}$ and $S_0=10.1$. Blue filled circles show values obtained during numerical integration of the equations of motion and the black solid curve is the analytical prediction from \cref{eq:freq_ratio}. The magenta curve is the analytical prediction from \cref{eq:freq_ratio} for $\alpha = \alpha_0$.}
    \end{center}
\end{figure}

\begin{figure}[t]
    \begin{center}
    \includegraphics[width=1\columnwidth,clip]{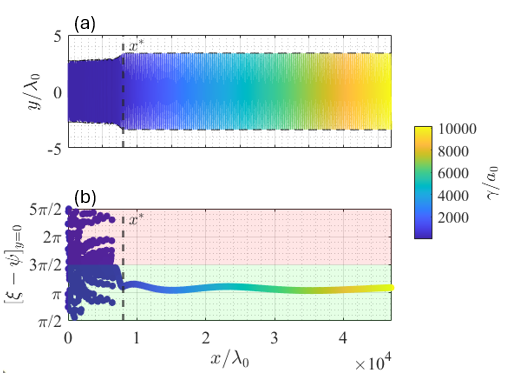}
    \caption{\label{fig:xStarPhase} The electron trajectory (a) and phase offset between the laser phase and the betatron phase (b) for an example from the scan in \cref{fig:kappaLinearXStarScan} with $\kappa=7.59 \times10^{-4}$ and $S_0=10.1$. In panel (a), the dashed curves show $\pm y_*$, where $y_*$ is the betatron amplitude given by \cref{eq:y_*}. In panel (b), the green shaded region indicates the favorable range and the red shaded region indicates the unfavorable range of the phase offset.}
    \end{center}
\end{figure}

To examine the underlying physics and make a clear comparison with \emph{Cases~1} and \emph{2}, we consider the parameters marked by the square symbol in \cref{fig:kappaLinearXStarScan}. They correspond to $S_0=10.1$ and $\kappa = 7.59 \times 10^{-4}$. Using the same $S_0$ as in \emph{Cases~1} and \emph{2} isolates the effect of the larger $\kappa$, which is higher by more than a factor of two. The $\gamma(x)$ profile in \cref{fig:xStarPlots} shows that intermittent energy losses are eliminated, so $\gamma$ increases without the characteristic peaks and reaches a much higher value over the same integration time. The phase offset in \cref{fig:xStarPhase} confirms that the electron remains within the favorable range throughout the evolution. The frequency ratio $\langle\omega'\rangle/\omega_{\beta}$ also remains stable (see \cref{fig:xStarFreq}), staying close to unity while $\gamma$ increases.

This regime can be understood using the analytical expression for $\langle\omega'\rangle / \omega_{\beta}$ given by \cref{eq:freq_ratio}. As discussed in \cref{sec-linear}, the second term inside the brackets dominates at sufficiently large $\gamma$ and the frequency ratio can be approximated by \cref{eq:freq_ratio-approx}, so that $\langle \omega' \rangle / \omega_{\beta} \propto \sqrt{\gamma/\alpha}$. Therefore, $\langle\omega'\rangle/\omega_{\beta}$ can remain stable if $\alpha$ increases at the same rate as $\gamma$. This is exactly what happens in the regime shown in \cref{fig:xStarFreq}.

The same stabilization was not achieved in \cref{sec-linear} because the large value of $\kappa$ required for it makes it difficult for the electron to reach the second resonance crossing in the first place. This failure mode was illustrated by \emph{Case~2}. As seen in \cref{fig:OvershootFreq}, the downward push of $\langle\omega'\rangle/\omega_{\beta}$ after the first resonance crossing is already too strong in that case, preventing the electron from reaching the second crossing. The value of $\kappa$ in the present example is even higher. Delaying the onset of the longitudinal variation of $\alpha$ to the second resonance crossing is what makes the key difference.

Finally, it is worth pointing out that the electron trajectory in this regime does not experience any inflation even though the electron experiences a substantial increase of $\gamma$. As shown in \cref{sec-alphaConst}, the transverse electron motion is confined to $|y| \leq y_*$, with $y_*$ given by \cref{eq:y_*}. At constant $\alpha$, an increase of $\gamma$ is known to lead to trajectory inflation with $y_* \propto \gamma^{1/2}$~[\onlinecite{Wang2020}], which can be detrimental because the loss of transverse electron confinement would terminate electron energy gain. We clearly see similar inflation when $\alpha$ is increasing. As seen in the example shown in \cref{fig:RampPhase}, the trajectory expands with the increase of $\gamma$. The key feature of the considered regime is that $\gamma/\alpha$ remains approximately constant during the increase of $\gamma$. At sufficiently high $\gamma$, $y_*$ in \cref{eq:y_*} is proportional to $\gamma/\alpha$. As a result, $y_*$ remains approximately constant even though $\gamma$ increases because the increase of $\alpha$ provides the necessary compensation. Therefore, in addition to eliminating intermittent energy losses, this regime also prevents trajectory inflation.

\section{Summary and discussion} \label{sec: summary}

In this work, we performed a detailed analysis of DLA assisted by a longitudinally nonuniform azimuthal plasma magnetic field. To isolate the impact of the nonuniformity, we used a test-electron model in which the laser field and the plasma magnetic field are prescribed. We found that a slow longitudinal increase of the field strength qualitatively changes DLA by introducing hysteresis in the dependence of the frequency ratio $\langle\omega'\rangle/\omega_{\beta}$ on $\gamma$. As a result, this ratio becomes dependent on the prior evolution of the electron, which makes it possible to control the phase offset between the laser phase and the betatron phase and thereby control the energy gain process.

Our analysis has revealed several advantageous features of the hysteresis and the resulting phase control. First, the hysteresis makes the energy exchange less reversible, so the energy gain does not have to be followed by a deep energy loss. Instead, $\gamma$ evolves as a sequence of increasing peaks with intermittent drops that do not erase the accumulated energy gain. As a result, the attainable $\gamma$ can be much higher than in the longitudinally uniform case and, in fact, we observe no saturation in our calculations. Second, the hysteresis transfers the electron onto a high-$\gamma$ trajectory that is not accessible during DLA at constant $\alpha$ when starting from low $\gamma$. This feature enables energy retention when the longitudinal variation of the magnetic-field strength is turned off because the electron cannot return to low $\gamma$. Finally, we find that the timing of the increase of $\alpha$ matters. If the increase begins after the electron has already reached high $\gamma$ in a longitudinally uniform field, then it becomes possible to enter a regime without intermittent losses in which $\gamma$ continues to increase.

In the present work, we intentionally neglected the photon emission and the corresponding recoil experienced by the electron. At the considered normalized laser amplitude of $a_0 = 8$, this is typically a reasonable step. There are DLA studies where test-electron models include this effect via the force of radiation friction~\cite{gong.scirep.2019,jirka.njp.2020,yeh.njp.2021,tangtartharakul.njp.2025}, but all of these cases involve a much higher $a_0$. This, however, does not automatically mean that the inclusion of radiation friction would be inconsequential for the regimes considered in our work. Previous work for a longitudinally uniform magnetic field has shown that radiation friction impacts the electron dynamics by changing the integral of motion $S$~\cite{yeh.njp.2021}. In that case, superluminosity was also shown to enhance the impact of radiation friction on DLA, which provides another potential connection to the present study~\cite{yeh.njp.2021}. Our regimes similarly rely on changes in $S$, so radiation friction might become relevant as the electron energy continues to increase. Even though this is beyond the scope of the present work, an important open question is how radiation friction would modify the hysteresis in the presence of a longitudinally varying plasma magnetic field.



\section*{Acknowledgments}

This material is based upon work supported by the Department of Energy National Nuclear Security Administration under Award Number DE-NA0004203 and by the U.S. National Science Foundation under Award No. PHY-2512067.

\section*{References}
\bibliography{Bib}


\end{document}